# A SIMULATION OF THE FERMILAB MAIN INJECTOR DUAL POWER AMPLIFIER CAVITIES*

S. M. Stevenson[1,†]

[1]Fermi National Accelerator Laboratory, Batavia, USA

## Abstract

The Fermilab Main Injector accelerating cavities have sparking issues when they are run at voltages higher than those required by the PIP-II project. This is a problem Fermilab is working on as planning begins for the next upgrade to the accelerator complex. One of the methods being used to address the issue is the development of a CST Microwave Studio simulation to accurately model the PIP-II dual power amplifier cavities and identify which part(s) of the cavity is causing sparking to develop. The model will also be used to determine if changes to the cavity geometry may allow the cavity to be used at higher voltages before sparking occurs.

## INTRODUCTION

The Main Injector at Fermilab is the last stage of the accelerator, accelerating the beam of protons up to 120GeV before it is sent off to the experiments. In order to achieve higher beam power levels in the future, the radio frequency accelerating cavities need to be able to generate more voltage to give the beam. Currently the maximum achievable cavity voltage is limited by arcing in the cavity. It would be beneficial for future accelerator operations to investigate how to mitigate the arcing limit. This paper will discuss the general cavity function, the planned cavity upgrades, and the development of a simulation to model the current behavior of the cavities with the goal of increasing their maximum voltage output.

## CURRENT CAVITY COMPONENTS

The cavity has three main parts: the body, the two power amplifiers, and the two ferrite tuners.

The body of the cavity acts like two back-to-back folded quarter wave coaxial resonators. A side view of the cavity is shown in Fig. 1 and a cross-section of the cavity is shown in Fig. 2. The folded aspect of the cavity results in three main elements of the cavity geometry: the inner conductor, the intermediate cylinder, and the outer conductor. The region of the cavity inside the intermediate cylinder is kept under a vacuum level of roughly 5e-8 Torr, while the part of the cavity outside the intermediate cylinder is not under vacuum. The outer conductor is shaped like an octagon instead of a cylinder, which both helps maintain structural integrity as well as make it easier to mount external components on the cavity. The intermediate cylinder and inner conductors are both cylindrical. The main feature of the intermediate cylinder is the ceramic regions on either end to allow fields to travel uninhibited through the cavity while the vacuum is preserved. The inner conductor's main feature is the two accelerating gaps in the center of the cavity. Beam passing through the cavity will be insulated from fields by the beam pipe until it passes across the two gaps before exiting the cavity.

The power amplifiers use tetrode style vacuum tubes to control the final power input into the cavity. There are two power amplifiers on each cavity that operate in a push-pull configuration. Each power amplifier is currently configured to operate with a maximum output of 150 kW. The power amplifiers are physically mounted on the top of the cavity and use a coupling loop to transfer power to the cavity with a 1:12.1 anode gap to accelerating gap voltage ratio, nominally generating a voltage of 210 kV across the gaps.

The two ferrite tuners are magnetically coupled to the cavity on the two lower 45 degree faces of the outer conductor. The ferrite tuners are used to tune the cavity from 52.808 to 53.104 MHz during the 1.2 seconds each batch of beam is in the Main Injector. Physically, they function as coaxial structures filled with alternating copper and ferrite disks. Each tuner is wound with 10 bus bar turns that act as an inductive structure to generate a magnetic field when a bias current is applied (note the bus bar turns are not shown in the model). This bias changes the ferrite disk permeability, causing the resonant frequency of the cavity to change, or tune.

## CURRENT VS. PLANNED CAVITY UPGRADES

Fermilab is currently executing an accelerator complex upgrade referred to as "PIP-II". Part of the PIP-II project is the addition of a second power amplifier to the existing Main Injector cavities in order to increase the power delivered to the beam. With the PIP-II project about half-way completed, Fermilab is now planning what the next goals of the accelerator should be, with a continuing emphasis on increasing the power of the proton beam. The method chosen to increase the power of the beam is to accelerate the same number of protons faster, with a goal of halving the time each batch of protons spends in the Main Injector. This necessitates a higher cumulative accelerating voltage from the cavities.

To achieve the greater voltage requirements, either the number of Main Injector accelerating cavities needs to be doubled or the amount of voltage each cavity can produce needs to be substantially increased. Building and installing that many more cavities is challenging due to cost, the civil construction required, and a lack of available space in the tunnel. This leaves generating more voltage from the current cavities, designing new ones that can produce the extra voltage needed, or some combination of the two. The table below lists the differences in functional requirements between PIP-II and the next generation of the accelerator.

Table 1: PIP-II vs. future accelerator requirements.

|  | PIP-II | Future |
|---|---|---|
| Acceleration Ramp Slope | 240 GeV/s | 500 GeV/s |
| MI Ramp Rate | 1.2 s | 0.65 s |
| 120GeV Beam Power | 1.2 MW | 2.22 MW |
| Beam Accelerating Power | 2.88 MW | 6 MW |
| Required Accelerating Voltage: $V\sin\phi_s$ | 2.66 MV | 5.54 MV |
| Total Voltage Available | 4.7 MV | 8.9 MV |
| Total Operating Voltage | 4.2 MV | 7.8 MV |
| Total Apparent Power | 240.5 kW/Cavity | 246.2 kW/Cavity |

Again, the main concern with using the current cavities to generate the necessary voltage levels is the fact that at higher voltages the cavities are known to start sparking. The exact cause and location of the sparking is not known, which is why a simulation has been built to identify where the sparking is likely to occur and if there is a geometrical solution that allows the cavities to be pushed harder before sparking occurs.

## SIMULATION AND RESULTS

The simulation was built in CST Microwave Studio based off of drawings made for the original 1973 Main Ring cavities. Each component was modeled and added to the simulation by its drawing number. The process was complicated by partially incomplete documentation and drawings for various parts of the cavities, as well as the fact that there are at least three different versions of the cavities installed in the tunnel. The final simulation used a combination of drawings from all three cavity versions as well as educated interpretation to simulate a cavity that meets the overall operating specifications of all the current Main Injector cavities. Comparison with data gathered from the real-life cavities was used to evaluate the simulation accuracy.

Both an Eigenmode solution and a Frequency Domain solution were run for the simulation. The Eigenmode solution generates the Q, R/Q, and gap voltage to be measured as well as showing the electric field distribution. The Frequency Domain solution was run to check the simulations S parameters from 40 MHz – 70 MHz, which were used to verify the frequency response of the cavity and check the cavity's tuning range.

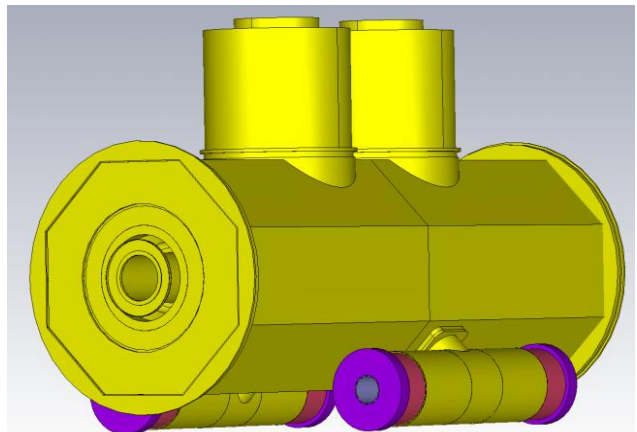

Figure 1: Side view of simulated cavity geometry.

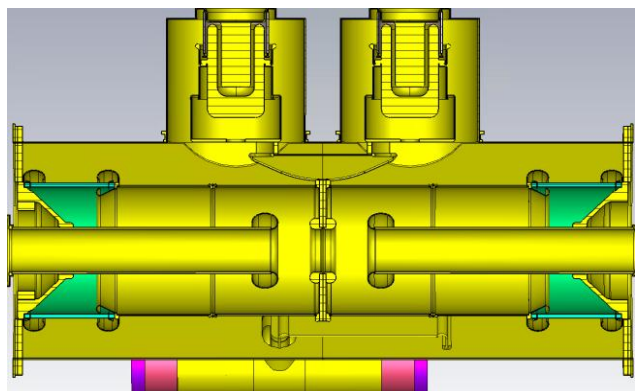

Figure 2: Simulated cavity geometry cross section.

The current unloaded cavity tuning range is verified by biasing the ferrite tuners from 41 to 315 amps, which should cause the cavity to tune from 52.808 to 53.104 MHz, a tuning range of 296 kHz. The simulation models the response of the ferrite to different bias currents via equations for the loss tangent and permeability of the material. The equations were formulated from physical measurements of a fully assembled tuner's response to bias current sweeps [1]. The simulation shows a frequency sweep from 52.723 to 53.172 MHz, a tuning range of 445 kHz from a current bias of 41 to 315 amps. While being somewhat larger than the measured tuning range, this result is still well within reason and proves that the simulated tuners are placing an adequate load upon the cavity. The frequency response of the cavity across the tuning range is shown in Fig. 3. This plot was generated by modelling weakly coupled coaxial probes on either side of the power amplifiers to measure the simulation's S21.

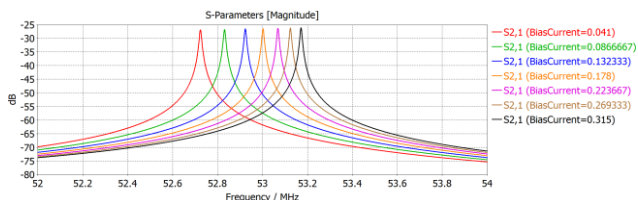

Figure 3: Cavity S21 parameters over the tuning range.

The nominal voltage across the gap of the cavity is 240 kV. The CST Eigenmode simulation generates a gap voltage of 278 kV, as this is the voltage associated with 1 Joule stored energy within the structure, which is the standard the program follows. Output values like Q, R/Q, and tuning range are unaffected by this, but field values require a scaling factor to convert to real world values. The scaling factor was chosen to be 86.33% to make the simulated gap voltage equivalent to the actual cavity voltage.

With the tuners attached, the cavities have a recorded Q of 5800. [2] The simulation shows a Q of 5892, which is well within reason.

When the gap is perturbed, the cavities have been determined to have a R/Q of 104. The simulation shows a R/Q of 115, which is slightly high but still acceptable. This difference can be accounted for by the fact that the simulation only includes the cavity, power amplifiers, and tuners, but not any of the higher-order mode dampers or other monitoring devices that place small loads upon the cavity.

The table below sums up the current cavity parameters vs. the simulation results.

Table 2 : Cavity Parameters

|  | Current Cavity | Simulation |
|---|---|---|
| Tuning Range | 296 kHz | 445 kHz |
| Gap Voltage | 240 kV | 240 kV |
| Q | 5800 | 5,892 |
| R/Q | 104 | 115 |

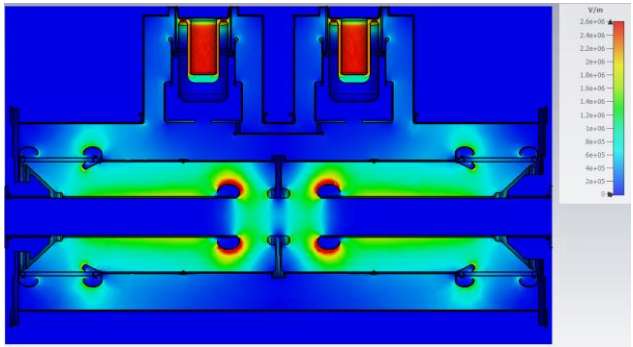

Figure 4: Electric Fields over the cavity cross section.

Fig. 4 illustrates the electric field distribution on a cross-section of the simulated cavity. The simulation shows that the highest electric fields in the cavity are located on the corona rings around the end of the beam pipes by the two gaps. The fields here have values up to 3.099E6 Volts/m vs. the 0.898E6 Volts/m across the gap. This suggests that the sparking is likely occurring on the inner conductor corona rings. The figure also shows high fields in the power amplifier tetrodes, however this is due to how the tetrode is simulated and is not a region of concern.

## CONCLUSION

A simulation has been developed that accurately models the current behavior of the Main Injector dual power amplifier cavities. The simulation shows that at larger voltages there are very high electric fields around the beam pipe corona rings that are likely responsible for some of the sparking the cavity experiences. Future work will focus on attempting to lower the peak electric fields in the cavity by changing the geometrical properties of the beam pipe corona rings. The center frequency, Q, and R/Q will be used as metrics to monitor the success of each modification.

Sincere thanks to Dr. Brian Vaughn and Dr. Danillo Erricolo for all their advice and help during this project.